\documentclass[final]{aipproc}

\layoutstyle{6x9}
\usepackage[english]{babel}
\usepackage{amsmath}
\usepackage{amssymb}
\usepackage{epsfig}
\usepackage{graphics,psfrag,rotating}
\usepackage{graphicx}

\begin{document}

\title{Indirect constraints to branon dark matter
}

\classification{11.10.Kk, 12.60.-i, 95.35.+d, 98.80.Cq}
\keywords{branons, dark matter, gamma rays}

\author{J.\,A.\,R.\,Cembranos}{
  address={Departamento de F\'{\i}sica Te\'orica I, Universidad Complutense de Madrid, E-28040 Madrid, Spain.}
}

\author{A.\,de la Cruz-Dombriz}{
  address={Astrophysics, Cosmology and Gravity Centre (ACGC), University of Cape Town, Rondebosch, 7701, South Africa.}
 ,altaddress={ Department of Mathematics and Applied Mathematics, University of Cape Town, 7701 Rondebosch, Cape Town, South Africa.}
}

\author{V. Gammaldi}{
address={Departamento de F\'{\i}sica Te\'orica I, Universidad Complutense de Madrid, E-28040 Madrid, Spain.} 
}

\author{A. L. Maroto}{
address={Departamento de F\'{\i}sica Te\'orica I, Universidad Complutense de Madrid, E-28040 Madrid, Spain.} 
}

\begin{abstract}
If the present dark matter in the Universe annihilates into Standard Model
particles, it must contribute to the gamma ray 
fluxes detected on the Earth. Here we briefly 
review the present constraints for the detection of gamma ray photons produced in the
 annihilation of branon dark matter. We show that observations of  
dwarf spheroidal galaxies and the galactic center
by  EGRET, Fermi-LAT or  MAGIC  are below
the sensitivity limits for branon detection. However, future experiments such as CTA
could be able to detect gamma-ray photons from annihilating branons of masses above 150 GeV.
\end{abstract}

\maketitle
%

\section{Introduction}

According to present astrophysical observations and collider experiments, dark matter (DM) cannot be accommodated within the Standard Model (SM)
of elementary particles. In the framework of indirect detection of DM, gamma photons might be observed as products of DM annihilation into SM particles. The magnitude
of such a contribution depends on the particular DM candidate and the 
astrophysical target.
Brane fluctuations (branons) are massive and weakly interacting particles 
which appear as natural DM candidates in brane-world models \cite {CDM,BR,ACDM}. 
Even if branons are stable, they can annihilate by pairs into ordinary particles in different astrophysical objects (galactic haloes, Sun, Earth, etc.) and generate cascade processes whose products could contribute to the gamma ray flux.

\begin{figure}[t]
\includegraphics[height=.37\textheight]{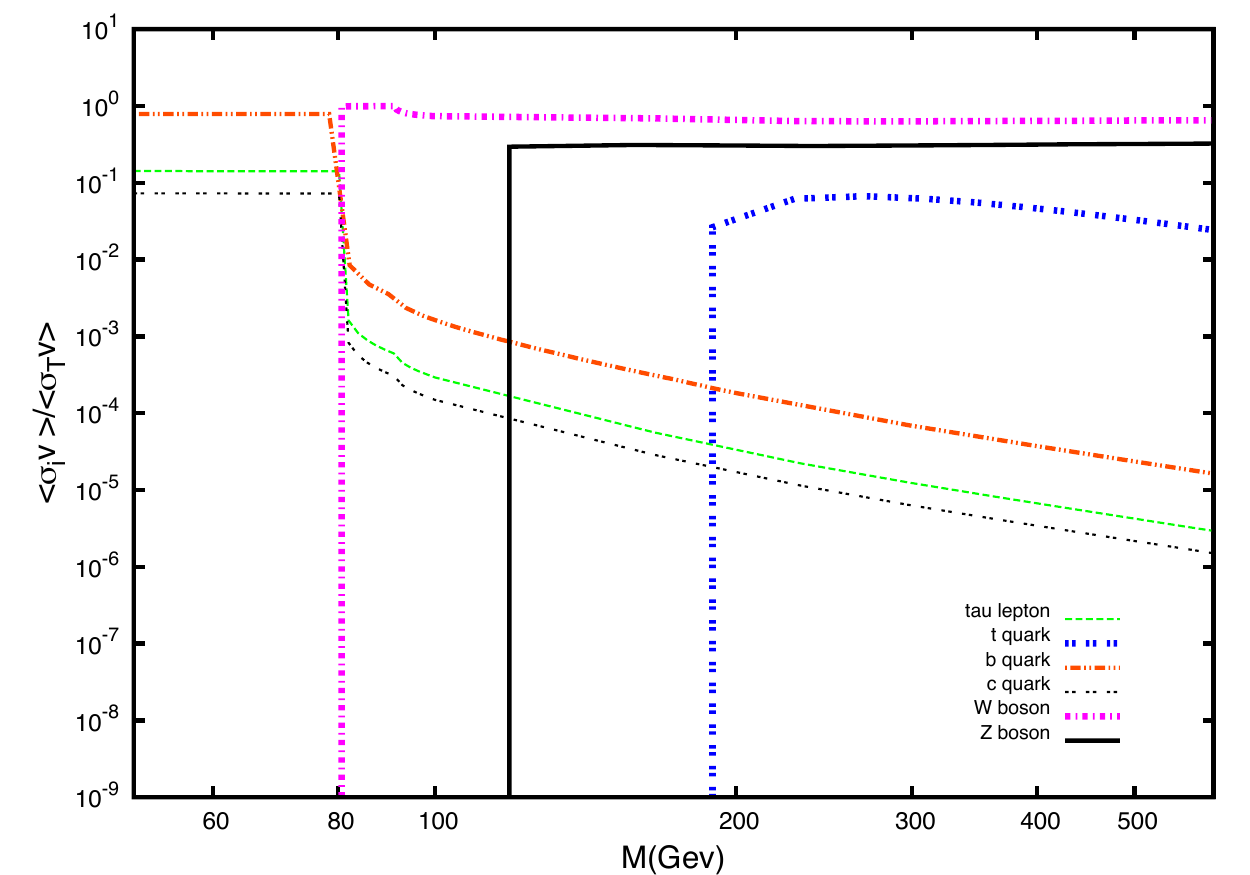}
\caption {\footnotesize{Branon annihilation branching ratios into SM particles. See text and Ref.\cite{branongamma}} }
\label{BR}
\end{figure}

This differential gamma ray flux from annihilating DM particles in galactic sources can be written as \cite{Ce10}:
\begin{eqnarray}
\frac{\text{d}\,\Phi_{\gamma}^{\text{DM}}}{\text{d}\,E_{\gamma}} =
\frac{1}{4\pi M^2}\sum_i\langle\sigma_i v\rangle
\frac{\text{d}\,N_\gamma^i}{\text{d}\,E_{\gamma}}\, \times\, \frac{1}{\Delta\Omega}\int_{\Delta\Omega}\text{d}\Omega\int_{l.o.s.} \rho^2[(s)] \text{d}s
\label{flux}
\end{eqnarray}
where the second term on the r.h.s. of this equation represents the astrophysical factor and the first term is the particle dependent part, with
$\langle\sigma_i v\rangle$ the thermal averaged
annihilation cross-section of two DM particles into two SM
particles (labeled by the subindex $i$). The number of photons produced in each decaying channel
per energy interval
$\text{d}\,N_\gamma^i/\text{d}\,E_{\gamma}$
can be simulated by means of the PYTHIA 
particle physics software \cite{Ce10}.
In the case of heavy branons, the main contribution to the photon flux comes from branon annihilation into
$ZZ$ and $W^+ W^-$ (Fig. \ref{BR}). In this case, the produced high-energy gamma photons could be in the range
(30 GeV-10 TeV), detectable by Atmospheric
Cherenkov Telescopes (ACTs)  such as MAGIC (with a threshold of 60-70 GeV). On the contrary, if $M<m_{W,Z}$, the
annihilation into $W$ or $Z$ bosons is kinematically forbidden and it is
necessary to take into account the rest of  channels, mainly
annihilation into the heaviest possible quarks (Fig. \ref{BR}).
In this case, the photon fluxes would be in the range detectable by
space-based gamma ray observatories 
such as EGRET (Energy range of 0.02-30 GeV) and FERMI (20 MeV-300 GeV).

%
%

\begin{figure}[t]
\includegraphics[height=.32\textheight]{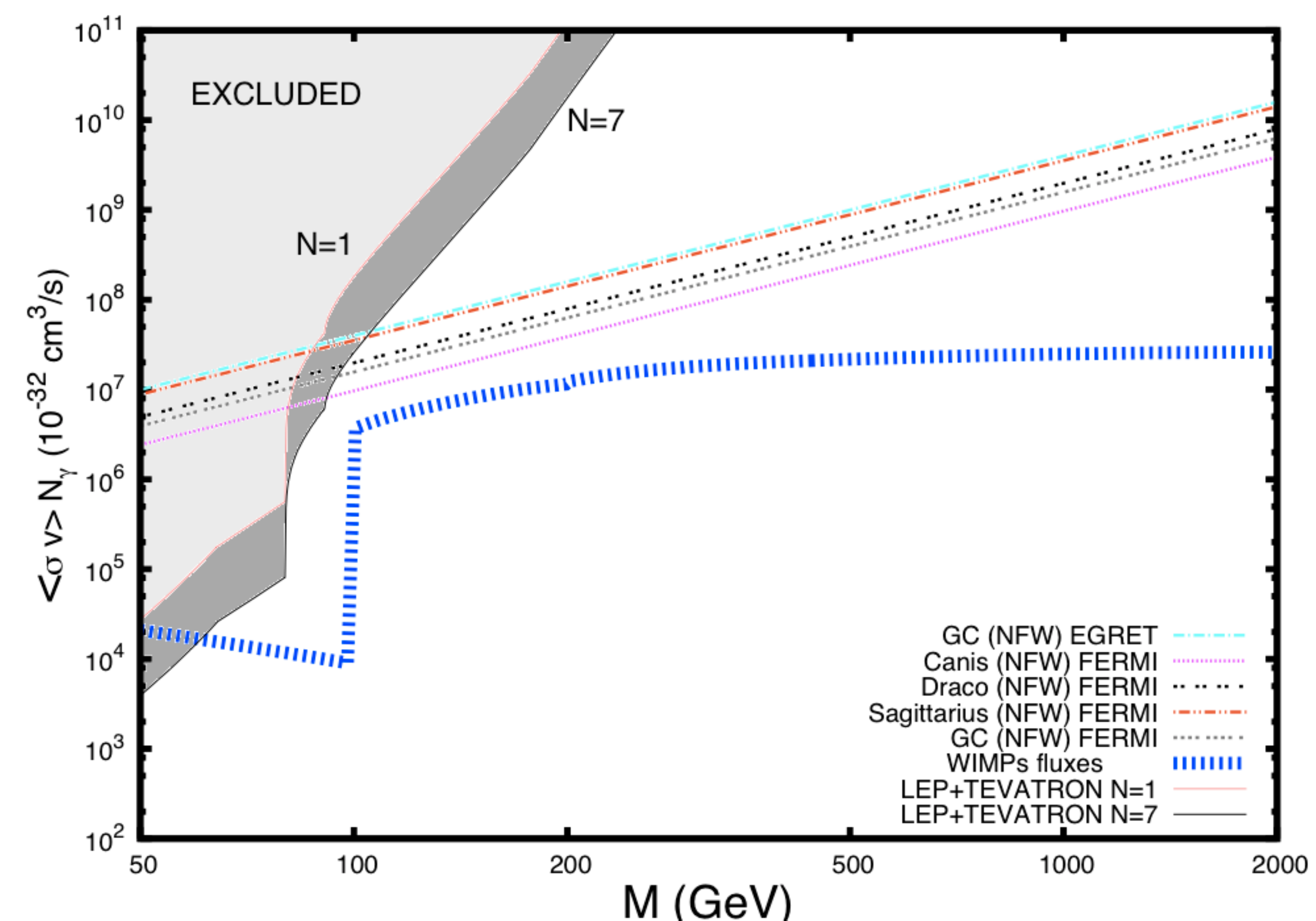}
\caption {\footnotesize Sensitivity of different targets to constrain gamma rays coming from branon annihilation. The straight lines
show the estimated exclusion limits at $5\sigma$ for satellite experiments (FERMI and EGRET).
The thick dashed line corresponds to the photon flux above 1 GeV coming from branons with the thermal abundance inside the WMAP7 
limits ($\Omega_{\text{CDM}} h^2 = 0.1123\pm0.0035$).
The area on the upper left corner above the corresponding lines is excluded by $\text{LEP}$ and $\text{TEVATRON}$ experiments for both $N = 1$ and $N=7$, number of extra dimensions \cite{ACDM}.}
\label{FER}
\end{figure}

\begin{figure}[t]
\includegraphics[height=.32\textheight]{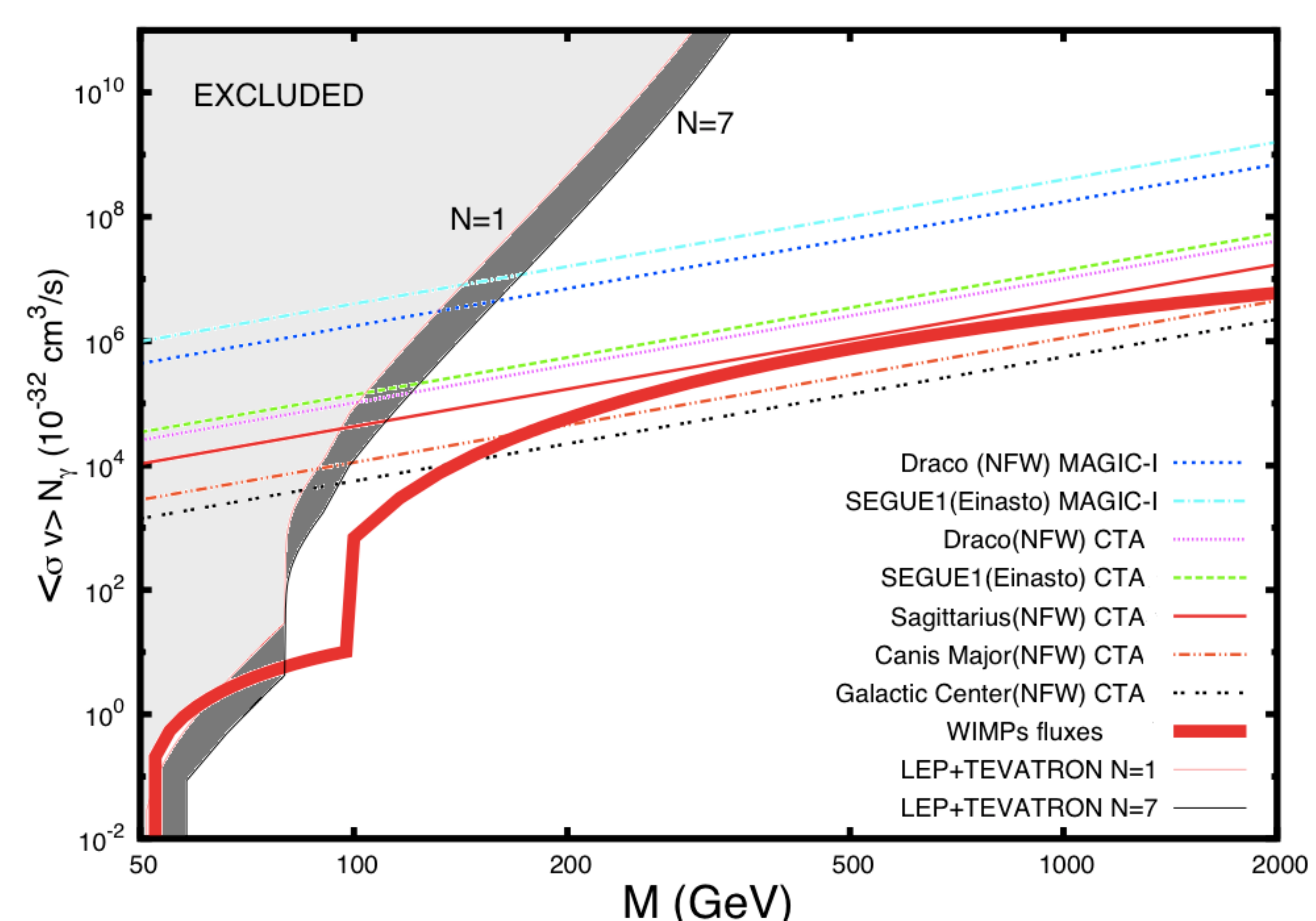}
\caption {\footnotesize Same as Fig. \ref{FER} for ground-based detectors. In this figure, the continuous thick dashed line
corresponds to the photon flux above 50 GeV coming from branons with the thermal abundance inside the WMAP7 limits.}
\label{ACT}
\end{figure}

\section{Analysis and results}

The best targets to search for a DM annihilation signal are dwarf spheroidals (dSphs) and the Galactic center. The astrophysical part $\langle J \rangle_{\Delta\Omega}$ of the gamma-ray flux (\ref{flux}) of each target depends upon the DM density. A Navarro-Frenk-White (NFW) profile is assumed for the Draco, Sagittarius and Canis Major dSphs and for the Galactic Center \cite{Ev04}, and an Einasto profile for SEGUE 1. 
The minimum  flux $\Phi_\gamma$ required for a detection with a significance above  $5\sigma$
can be obtained from \cite{Ev04}:
\begin{equation}
\frac{\Phi_\gamma\sqrt{\Delta\Omega\,A_{eff}\,t_{exp}}}{\sqrt{\Phi_\gamma+\Phi_{Bg}}} \geq 5\; .
\label{minflu}
\end{equation}
where $t_{exp}$ is the exposure time,    $A_{eff}$  the instrument  effective area 
and $\Delta\Omega$ the angular acceptance. 
The evaluation of the background $\Phi_{Bg}$ and its value depends both on the experiment and on the source. 
%
In \cite{branongamma}  the astrophysical factor, the technical details of each experiment, the background estimations and the resulting values of the minimum detectable gamma-ray fluxes for each sources are reported.

By using the estimated minimum detectable flux at 5$\sigma$ significance and the particular astrophysical factor ($J_{\langle\Delta\Omega\rangle}$) of each target, the sensitivity on $N_\gamma^{(5)}
\langle\sigma v\rangle$   has been obtained as a function of the WIMP mass depending on the particular detector.  The corresponding curves for the different targets and
detectors are shown in Figs. 2 and 3. The theoretical value for  $N_\gamma
\langle\sigma v\rangle$ for branons has been obtained by integrating the differential
spectrum $\sum_i\langle\sigma_i v\rangle
\frac{\text{d}\,N_\gamma^i}{\text{d}\,E_{\gamma}}$ taking into account the energy threshold
of 1 GeV for satellite experiments (Fig. 2) or 50 GeV for ACTs (Fig. 3). The resulting  $N_\gamma
\langle\sigma v\rangle$ with the WMAP constraints on the relic density, is a function of M \cite{branongamma}. 

\section{Conclusions}
As shown in Fig. \ref{FER},  present experiments (EGRET, FERMI or MAGIC) are unable to
detect signals from branon annihilation for the targets considered. However, as shown in
Fig. 3, future experiments such as CTA
could be able to detect gamma-ray photons coming  from the annihilation of branons with masses
higher than 150 GeV for observations of the Galactic Center or above 200 GeV for Canis Major.
In the same figures (\ref{FER} and \ref{ACT}), it is possible to see the present constraints from collider experiments. These searches
are complementary and probe, in general, a different area of the parameter space of the model.


\begin{theacknowledgments}
We would like to thank Daniel Nieto for useful comments. This work has been supported by MICINN (Spain) project numbers FIS 2008-01323, FIS2011-23000, FPA 2008-00592, FPA2011-27853-01 and Consolider-Ingenio MULTIDARK CSD2009-00064. AdlCD also acknowledges the URC (University Research Council) and the National Research Foundation (South Africa). 
\end{theacknowledgments}

\bibliographystyle{aipproc}   

\end{document}